Title: Chromospheric mass motions and intrinsic sunspot rotations for NOAA Active Regions 10484, 10486, and 10488 using ISOON data

Short title: Dynamical motions within three active regions

Paul S. Hardersen
Department of Space Studies, University of North Dakota, 4149 University Avenue, 530 Clifford Hall, Grand Forks, ND 58202-9008, Hardersen@space.edu

K.S. Balasubramaniam
Solar and Solar Wind Disturbances Program, Space Vehicles Directorate, Air Force Research Laboratory, Kirtland AFB, NM 87117

Svetlana Shkolyar
School of Earth and Space Exploration, Arizona State University, 781 Terrace Road, ISTB4, Tempe, AZ 85287-6004

Abstract. This work utilizes Improved Solar Observing Optical Network (ISOON: Neidig et al. 2003) continuum (630.2 nm) and Hα (656.2 nm) data to: 1) detect and measure intrinsic sunspot rotations occurring in the photosphere and chromosphere, 2) identify and measure chromospheric filament mass motions, and 3) assess any large-scale photospheric and chromospheric mass couplings. Significant results from October 27-29, 2003, using the techniques of Brown et al. (2003), indicate significant counter-rotation between the two large sunspots in NOAA AR 10486 on October 29, as well as discrete filament mass motions in NOAA AR 10484 on October 27 that appears associated with at least one C-class solar flare.

1. Introduction

Solar atmospheric activity has often been suggested as a precursor to solar flares, which can assist in devising techniques to accurately predict space weather events. Intrinsic sunspot rotations have been detected for decades (Evershed 1910; St. John 1913; Abetti 1932; Maltby 1964; Gopasyuk 1965; Bhatnagar 1967; McIntosh 1981; Pevtsov & Sattarov 1985; Ding et al. 1987), but their association with solar flare generation is uncertain. Filament activity prior to, and during, flaring also occurs frequently (Jiang and Wang 2001; Jing et al. 2004; Yan et al. 2012), but establishing diagnostic pre-flare indicators remains elusive.

A history of detecting and measuring mass motions in the solar atmosphere exists, which includes intrinsic sunspot rotations and filament motions across the solar disk. Intrinsic sunspot rotation, defined as proper rotation of a sunspot around an axis perpendicular to the solar surface and centered on the umbra, has been detected since the early 20$^{th}$ century utilizing both ground- and space-based photospheric and magnetogram data (Evershed 1910; St. John 1913; Abetti 1932; Maltby 1964; Gopasyuk 1965; Bhatnagar 1967; McIntosh 1981; Pevtsov & Sattarov 1985; Brown et al. 2003; Min & Chae 2009; Yan & Qu 2007; Zhang et al. 2007; Yan et al. 2008, 2009; Kazachenko et al. 2009, 2010).

Recent years have seen additional work on this topic thanks to spacecraft such as *SOHO* (Domingo et al. 1995), *TRACE* (Schrijver et al. 1996), and *Hinode* (Kosugi et al. 2007), along with ground-based observatories such as the Improved Solar Observing Optical Network (*ISOON*) (Neidig et al. 2003).

Despite abundant detections of rotating sunspots in the past century, the causes and implications of this phenomenon are uncertain. Yan et al. (2008) conducted a statistical study of 182 measured sunspot rotations using *SOHO*, *TRACE*, and *Hinode* data to constrain sunspot rotational properties. While Yan et al. (2008) report that there is no rotational [clockwise (CW) vs. counterclockwise (CCW)] preference in the northern and southern solar hemispheres, the northern hemisphere has a greater abundance (12%) of rotating sunspots while positive (negative) polarity in the northern (southern) hemisphere is about twice as common for rotating sunspots.

In a statistical study of chromospheric penumbral whorls as traced by superpenumbral filament directions, Balasubramaniam et al. (2004) found only a weakly correlated sense of twist of penumbral fields for following hemispheric preference. However, their data were not intended to study the temporal evolution of these twists as related to flares.

Intrinsically rotating sunspots appear to be relatively rare occurrences as only ~5% of the analyzed active regions in Yan et al. (2008) exhibited rotations. The occurrence of more than one rotating sunspot per active region is more rare (~1%) with only a minor preference for multiple sunspots to rotate in the same direction as compared to opposite directions (Yan et al. 2008).

Multiple causes for these rotations have been proposed, but a community consensus on the most dominant, or sole, mechanism is still lacking. Brown et al. (2003) suggests photospheric flows and magnetic flux tube emergence as potential mechanisms, while Yan & Qu (2007) suggest a combination of the Coriolis force and differential rotation as alternative mechanisms. Su et al. (2008) investigated the interactions of subsurface vortical flows and magnetic flux tubes as possible drivers of sunspot rotation.

Investigating the chromospheric surface above sunspots, Balasubramaniam et al. (2004) have shown that the sunspot magnetic structure is preserved above the photospheric layer, hence, rotations should be well correlated between photosphere and chromosphere.

Brown et al. (2003) suggests that measured sunspot rotation rates are orders of magnitude greater than what would be expected from rotations caused by differential rotation or the Coriolis force. Table 1 shows a sampling of the average or maximum intrinsic sunspot rotation rates reported in the literature. Note that most reports of intrinsic sunspot rotations display variable rotation rates through time when a sunspot is rotating (Brown et al. 2003; Min & Chae 2009).

Table 1. Representative intrinsic sunspot rotation rates reported in the literature. Average or maximum rotation rates are reported. Note that rotation rates are usually variable through time when rotation occurs. Rotational velocities calculated, when possible, using Equation 1 from Brown et al. (2003).

| NOAA Active Region | Peak or Average Rotation Rate (° hr$^{-1}$) | Peak or Average Rotational Velocity (km s$^{-1}$) | Rotational Direction | Reference |
|---|---|---|---|---|
| 10759 | 0.85 (avg.) | -- | CCW | Kazachenko et al. (2009) |
| 10930 | 8 (peak) | -- | CCW | Min and Chae (2009) |
| 10930 | 20 (peak) | -- | CCW | Zhang et al. (2007) |
| 8668 | 1.3 (peak) | 0.055 | CCW | Brown et al. (2003) |
| 9004 | 3 (peak) | 0.095 | CW | Brown et al. (2003) |
| 9077 | 1.2 (peak) | 0.030 | CCW | Brown et al. (2003) |
| 9114 | 2.2 (peak) | 0.077 | CCW | Brown et al. (2003) |
| 9280 | 0.8 (peak) | 0.031 | CCW | Brown et al. (2003) |
| 9354 | 1.4 (peak) | 0.030 | CCW | Brown et al. (2003) |
| 10030 | 2 (peak) | 0.077 | CCW | Brown et al. (2003) |
| 10424 | 2.7 (avg.) | -- | CCW | Yan and Qu (2007) |
| 8210 | few degrees/hour | -- | CW | Regnier and Canfield (2006) |
| 10486 | 1.2 (leading spot, avg.) | -- | CCW[1] | Zhang et al. (2008) |
| 10486 | 1.5 (following spot, avg.) | -- | CCW | Zhang et al. (2008) |
| 10486 | 1.5 (leading spot, avg.) | 0.200 (+ 0.023/- 0.028) | CW | This work. |
| 10486 | 0.76 (following spot, avg.) | 0.131 (+ 0.011/-0.015) | CCW | This work. |

[1] For the leading sunspot in NOAA AR 10486, Zhang et al. (2008) report counter-clockwise rotation on five of the six days the sunspot was measured. Small clockwise rotation was measured on 10/27/2003. On 10/29/2003, Zhang et al. (2008) report that the leading sunspot rotated at an average angular speed of 4.38° hr$^{-1}$.

Another question involves the relationship between intrinsic sunspot rotations and the occurrence of flares. If a causative connection exists, then it may be possible to predict flaring, which would have implications for space weather prediction efforts. Stenflo (1969) and Barnes & Sturrock (1972) suggest that intrinsic sunspot rotation can lead to magnetic energy buildup and release, directly followed by a solar flare. Brown et al. (2003) studied seven active regions and compared time-coincident occurrences of sunspot rotations and flares. While flares occurred in six of the seven active regions during sunspot rotations, Brown et al. (2003) suggests that only three of the flares are likely to be associated with sunspot rotations. Min & Chae (2009) report significant rotation of the small sunspot in NOAA AR 10930, but no flaring has been causatively linked with this rotation.

Yan & Qu (2007), Zhang et al. (2007), and Yan et al. (2009) suggest direct relationships between rotating sunspots and flares while Kazachenko et al. (2009) suggests that sunspot rotation injected significant helicity into the corona of NOAA AR 10759 that led to flaring. Kazachenko et al. (2010), however, report that sunspot rotation in NOAA AR 10486 did not contribute significantly to the helicity and energy budgets that lead to some of the major solar flares during the 2003 'Halloween storms'. The summation of past research on connections between intrinsic sunspot rotation and flaring requires further demonstration and a larger, detailed effort to search for systematics in sunspot rotational behavior.

Chromospheric filaments are another atmospheric phenomena that have associations, and possible implications, to solar flares. Filament mass motion and flare initiation have been studied by many workers (Jing et al. 2003, 2004; Sterling & Moore 2005; Rubio da Costa et al. 2012; Vemareddy et al. 2012; Zuccarello et al. 2009). This research will add to the growing body of knowledge concerning both intrinsic sunspot rotation and filament mass motion, which often contributes to, or are associated with, solar flares.

2. <u>Observations and Methodology</u>

Continuum (630.32 nm) and Hα (656.28 nm) imagery for this research were obtained from the USAF Improved Solar Observing Optical Network (ISOON) on 2003 October 27, 28, and 29. At the time of the dataset, ISOON was located at the National Solar Observatory, Sacramento Peak, New Mexico, at an altitude of 2939 meters. Primary ISOON optical and mechanical components include a 250 mm diameter objective, independent right ascension and declination motor drives, tunable 0.1 Å Fabry-Perot etalons, and Cooke triplet lenses. CCD images were obtained by a cooled (-10°C) 2048 x 2048 pixel array with a plate scale of 1.08437 arc sec pixel$^{-1}$. Original ISOON solar images have a plate scale of 781 km pixel$^{-1}$ or 720 km arc sec$^{-1}$. The plate scale for flattened images is 498 km pixel$^{-1}$ or 459 km arc sec$^{-1}$. Dark current and flat field calibration frames were automatically applied to all images immediately following image acquisition. Figure 1 shows an example of original and flattened/limb-darkening-removed ISOON continuum and Hα images.

Continuum images on October 27 and 29 had a cadence of 30 minutes or 1 hour ranging from ~14:00-23:30 UT or within a subset of that timeframe. Sets of three continuum images were taken during each active observation with the highest-resolution image used for analysis. Only five sets of continuum images on October 28 were available, which did not allow analysis of photospheric motions for that day. Hα images with a 1-minute cadence were available from ~14:00-23:00 UT on all three days. Data gaps on October 29 from 17:22-18:01 UT, 19:46-20:03

UT, 20:13-21:01 UT, and 23:30-23:46 UT exist in the Hα dataset due to clouds. Hα image analysis used images with a 10-minute cadence.

NOAA Active Regions (AR) 10484, 10486, and 10488 were the most prominent active regions visible in the dataset and are the focus of this paper. Data analysis techniques applied to these active regions include sunspot uncurling to measure rotational motions in the photosphere and chromosphere, and detection of filament motion during active region evolution via IDL movie routines and SAOImage DS9 (Joye & Mandel 2003; Joye 2006).

Uncurling sunspots to measure rotational motion follows the techniques described in Brown et al. (2003). Using IDL routines, the ISOON images chosen for analysis were first flattened with limb darkening removed. In each set of continuum and Hα images, each active region was extracted by defining a large rectangular area around the active region. The cropped set of images for each active region were then aligned relative to a reference image, normalized, and trimmed to remove edge artifacts.

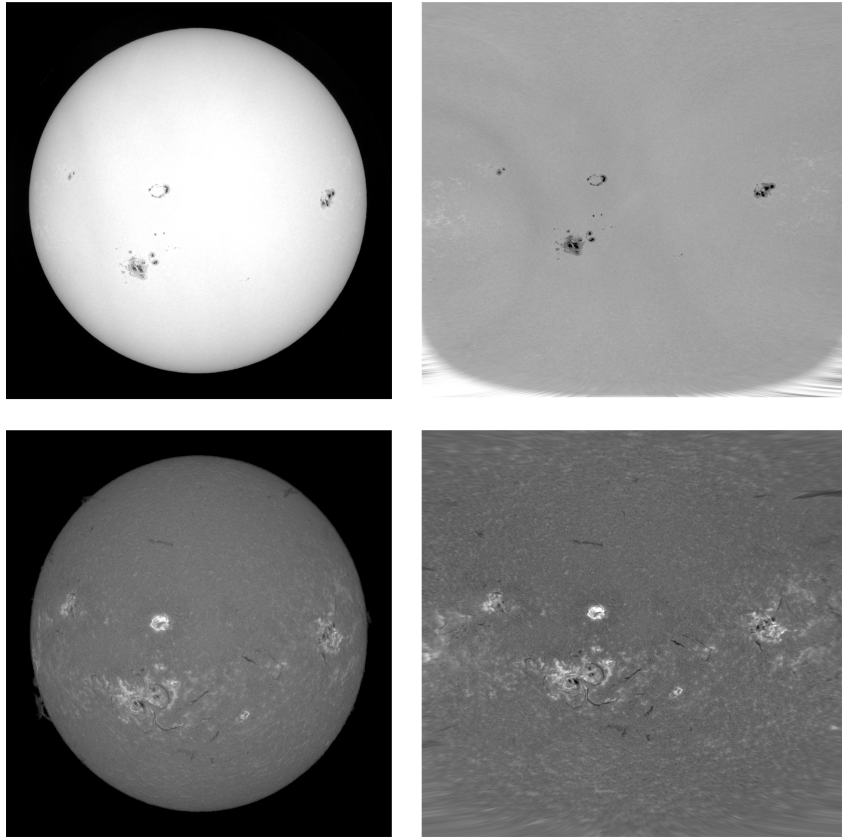

Figure 1. Representative ISOON continuum images (top row) and Hα images (bottom row) from October 27, 2003. Images at left are the original ISOON images. The images to the right represent the images at the left after being flattened with solar limb darkening removed.

Both large sunspots in NOAA AR 10486 and 10488 were uncurled, while the leading and trailing sunspots of the four-sunspot complex in NOAA AR 10484 were uncurled. For each sunspot, a circular aperture was centered on each sunspot of a chosen radius that extends well beyond the penumbra. Each sunspot is then uncurled – transformed from an *(x,y)* to *(r,ϑ)* coordinate system -- in 1° increments beginning with the eastern chord (0°) and moving sequentially counterclockwise. Figure 2 shows an uncurled image for the following sunspot of NOAA AR 10486.

Measuring rotational and dynamical motion of sunspots and filaments requires extracting radial time-slices from uncurled images and measuring non-horizontal motion in *(ϑ,t)* space [see Fig. 7, Brown et al. (2003)]. Features that exhibit a change in positional angle through time (Δθ/Δt) exhibit rotational motion while those not showing such angular movement do not. Four radial slices with 10-pixel-widths (10.84"), ranging from the inner penumbra to the outer penumbra/quiet photosphere, were extracted from each uncurled sunspot, aligned chronologically on each day, and examined for rotational motion [in *(ϑ,t)* space]. The pixel coordinates for each feature (continuum and Hα) were manually measured across the timespan of the data to determine its position, θ, which allowed calculation of feature rotation rate in θ hr$^{-1}$ and km s$^{-1}$ via Eq. (1) in Brown et al. (2003). Ten measurements were made for each time slice with the mean and standard deviation calculated to determine feature position through time. Figure 3 shows both a time-slice plot of the following sunspot rotation on October 29 for NOAA AR 10486 and the two features in that region that were measured.

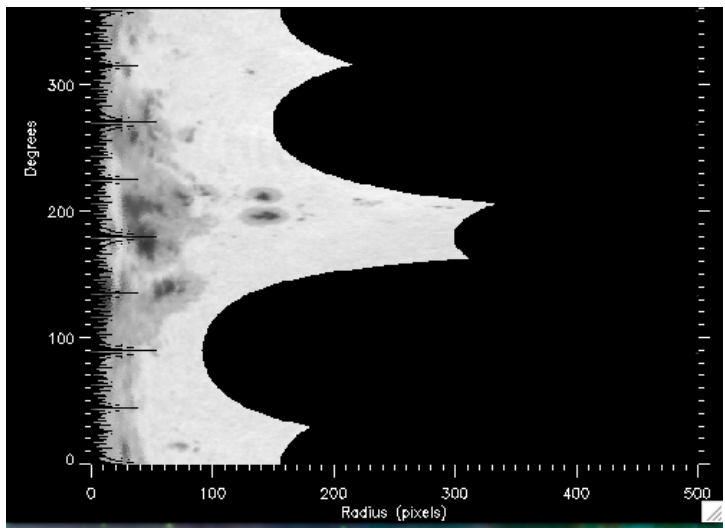

Figure 2. A representative uncurled continuum image (*ϑ,r*) of the following large sunspot in NOAA AR 10486 on October 29. The image shows the dark umbra stretched across the first ~15 radial pixels and the asymmetrical leading sunspot at ~180° along with its surrounding penumbra and pores. Some artifacts from the reduction process are present along the y-axis. See the text for details of the image uncurling process.

IDL movie routines and SAOImage DS9 were also used to study time-lapse movies of the three active regions. The primary benefit of this effort was to correlate chromospheric filament motions in each active region with the rotational motions seen in the Hα time-slice plots of the uncurled sunspots. Movies were also beneficial in detecting very fast and/or complex filament motion that is not detectable in the time-slice plots.

Positive and negative magnetic field areas in SOHO magnetogram images were measured through time using the IDL program, Helioflux (Panasenco et al. 2011). This was conducted to understand the broader magnetic field context in each active region through the time of the dataset.

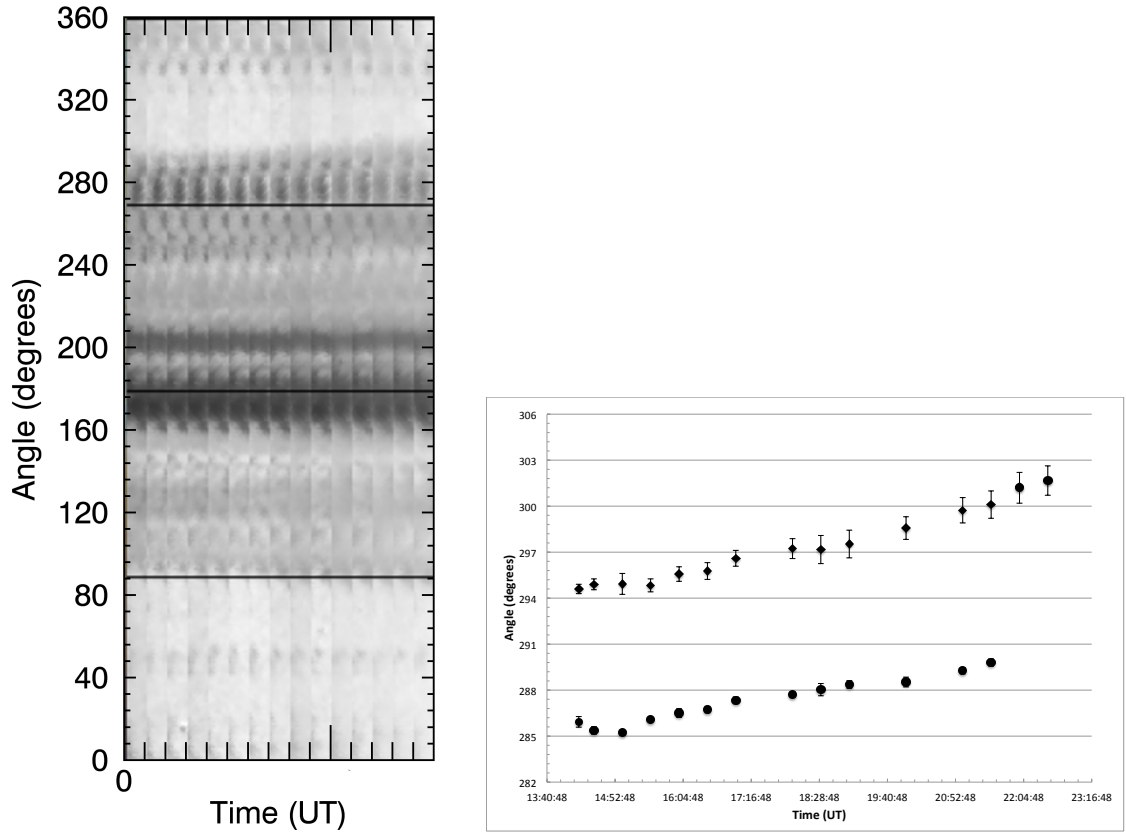

Figure 3. 3a (left). An angle vs. time plot for the following sunspot in NOAA AR 10486 on October 29. Time slices 10 pixels wide were extracted from continuum images and aligned chronologically. Note the wide area of rotational (i.e., non-horizontal) motion from ~230-300°. The time range for the plot is from ~14:15 UT to ~22:30 UT. 3b (right). Plot showing the measurement of two of the features within the area of rotation in Figure 3a. Error bars represent standard deviation from 10 measurements of the center of each feature.

3. Results

    3.1.   NOAA AR 10484

NOAA AR 10484 became visible on the solar surface on October 18 and remained visible until rotating out of view on October 31. At its peak, this AR had a βγδ magnetic classification

(http://solar.ifa.hawaii.edu/ARMaps). Continuum ISOON images from October 27 reveal an active region with five or six discernable umbrae of various sizes with penumbra linking the umbrae together. Hα images on October 27 clearly show the sunspots surrounded by diffuse chromospheric brightening around the active region, dominantly to the southwest of the sunspots. Visible filaments are present to the northeast, southeast, south, and northwest of the sunspots.

Two C-class solar flares occurred in our dataset near AR 10484 on October 27, including a C7.5 flare that began at 18:34 UT and a C9.0 flare that began at 19:48 UT (http://www.lmsal.com/solarsoft/last_events_20031028_0955/index.html). The C7.5 flare shows Hα brightening to the south of the leading sunspot and to the northeast of the sunspots while Hα brightening for the C9.0 flare occurred solely to the northeast of the active region.

No rotational motion was apparent in the continuum images on October 27 for either the leading or following sunspot. Two dark chromospheric features in time-slice sequences at average distances of 45 pixels (or 48.8") from the umbral center were seen in uncurled images of the following sunspot at ~145° and ~290°, respectively. The feature at ~145° displayed motion from ~16:33 UT to ~21:20 UT while the feature at ~290° showed rotational motion from ~16:33 UT to ~19:45 UT. Both features, which are likely associated with filaments, moved counterclockwise at average rates of 2.28° $hr^{-1}$ (0.39 km $s^{-1}$) and 1.10° $hr^{-1}$ (0.19 km $s^{-1}$).

Time-series analysis of AR 10484 Hα images in SAOImage DS9 (Joye and Mandel 2003; Joye 2006) show a series of discrete filament motions north/northeast of the following sunspot prior to both the C7.5 flare and the C9.0 flare. Filament material followed a reverse-S path that begins just north of the following sunspot at ~16:13 UT and moves in a distinct reverse-S pattern north of the same sunspot by ~17:43 UT. Figure 4 displays a sequence of images tracing the path of the filamentary mass, which can be seen traveling up to ~95" (~44,000 km) from the following sunspot (a movie of this motion is provided with the online version of this paper). This movement precedes the occurrence of the C7.5 flare by ~2 hours and is spatially coincident with brightening resulting from the flare.

A second burst of filament-like mass motion occurred from ~19:14 UT to ~19:24 UT in a direction that is halfway along the reverse-S path moving to the north (about the same position as the 17:13 UT image in Figure 4). The path of both sets of filamentary motion is spatially coincident and the second set of mass motion just precedes the C9.0 flare, where brightening occurs at ~19:48 UT. Filamentary mass motion occurred along the northwest boundary of the Hα brightening from the C9.0 flare prior to the occurrence of the flare.

On October 28, there was no detectable photospheric mass motion in continuum images. Hα chromospheric images showed filamentary material moving CCW at an average rotational speed of 12.8° $hr^{-1}$ (1.17 km $s^{-1}$) from ~14:30 UT to ~22:30 UT. The rotational rates were variable with somewhat faster rotation occurring from ~14:30 UT to ~18:40 UT and a slower rotational rate from ~18:40 UT to ~22:30 UT. Filament motion on October 28 must have occurred by a mechanism other than sunspot rotation due to the lack of any detectable rotation. Figure 5 displays the rotational characteristics of the filamentary motion on October 27 and 28.

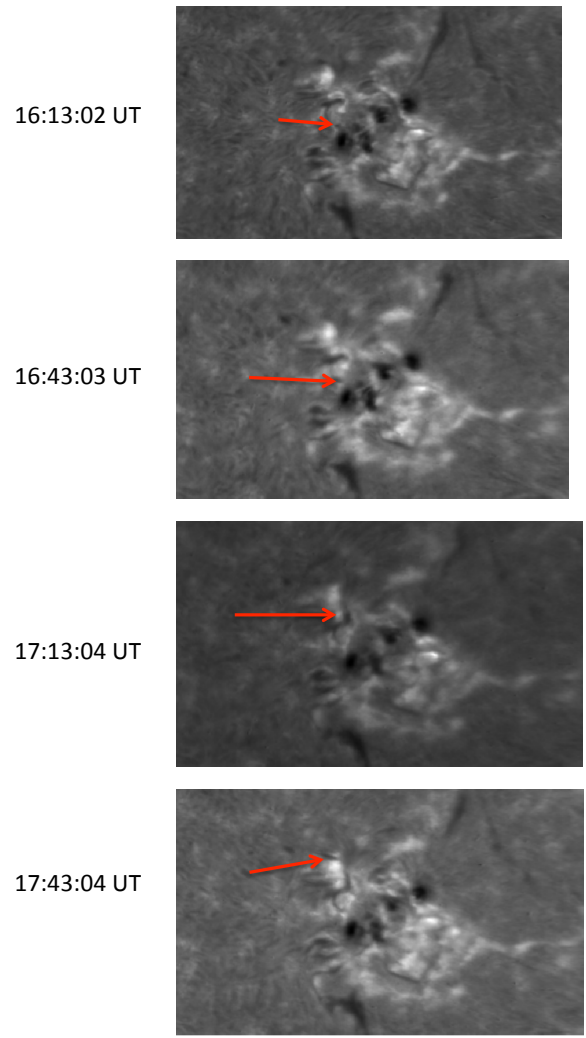

Figure 4. Time sequence of four Hα images of NOAA AR 10484 on October 27, 2003, from 16:13 UT to 17:43 UT. Discrete filament mass can be seen migrating north in a reverse sigmoid fashion, beginning with the top image and moving downward. Additional mass movement at ~19:30 UT directly precedes a C9.0 class flare that begins at ~19:44 UT. A movie of this filament-like motion is available in the online version of this paper.

Figure 5. Rotational motion of filaments as measured from time-slices of the following sunspot in NOAA AR 10484 on October 27, 2003, top, and on October 28, 2003, bottom. The motion on October 27, 2003, corresponds with filament mass motion moving to the north/northeast of the following sunspot that is time-coincident with a C9.0-class solar flare. Vertical bars in the top figure represent the approximate times of the C7.5 and C9.0 flare, respectively.

## 3.2. NOAA AR 10486

NOAA AR 10486 was large and active during the time range of the dataset and one of the principal contributors to the 'Halloween' solar storms in late October and early November 2003 (Hady 2009). AR 10486 emerged on the solar disk late on October 23 as a large, mature βγδ active region. A large X-class flare occurred during the three days of our data – an X10 flare on October 29 that began at ~20:37 UT. This flare occurred the day after an ~X17 flare that began at ~9:51 UT on October 28 that has been extensively studied (Ambastha 2007; Zuccarello et al. 2009; Kazachenko et al. 2010), but is not available in our data. SOHO magnetogram measurements of AR 10486 for the three days show increasing and decreasing negative and positive magnetic field areas, respectively, prior to October 27, and leveling off of the magnetic field areas early on October 28. Magnetic field areas of both polarities remain constant through the rest of the dataset with the negative magnetic field area ~1.3 times larger than the positive magnetic field area.

On October 29, significant photospheric and chromospheric mass motion occurs before, during, and after the X10 flare that is prominent in the Hα images. This motion is mostly consistent with, and a subset of, the five days of rotational motion seen in NOAA AR 10486, as reported by Zhang et al. (2008). Continuum images of the following sunspot show a wide swath of CCW photospheric mass motion from ~220° to ~300° (Figure 3a). Three distinct features within the overall area of motion were measured at an average distance of 48.8" (45 pixels) from the umbral center and show mass motion with rotational velocities of 0.70° hr$^{-1}$ (0.12 km s$^{-1}$), 0.71° hr$^{-1}$ (0.12 km s$^{-1}$), and 0.86° hr$^{-1}$ (0.15 km s$^{-1}$) (Figure 3b). Using an average photospheric density of 10$^{-9}$ g cm$^{-3}$, a vertical photosphere thickness of 500 km (Eddy, 1979), and gas moving across ~70°, the approximate mass of photospheric gas entrained in rotation from ~14:20 UT to ~22:30 UT was ~7 x 10$^{17}$ kg.

In contrast, continuum images of the leading sunspot show a much more restricted region of CW mass motion occurring from ~14:00 UT to ~21:00 UT with an average velocity of 1.50° hr$^{-1}$ (0.20 km s$^{-1}$). The feature was measured at an average distance of 38" (35 pixels) from the umbral center. This feature has been identified as a distended segment of the leading large sunspot's umbra, which extends to the west/southwest of the primary umbra. This motion is in the opposite direction to that reported by Zhang et al. (2008), but involves different areas of the sunspot.

Hα images of the following sunspot show systematic CCW motions (through ~3-4°) of bright regions in the ~14:20 UT to ~17:00 UT time frame at average rates of ~0.67° hr$^{-1}$ (0.14 km s$^{-1}$) and ~1.55° hr$^{-1}$ (0.33 km s$^{-1}$). Rotation of both features ceases after ~18:00 UT until the onset of the X10 flare at ~20:37 UT. Two features in the region of flare brightening of the following sunspot appear to shift abruptly CCW by ~8° and ~14° before returning to relatively constant, but displaced, positions. Hα images of the leading sunspot also show a CCW shift of material by ~10° before moving CW after the flare to return to within ~2° of the pre-flare position.

### 3.3. NOAA AR 10488

NOAA AR 10488 appears in SOHO magnetograms on October 26 and grew rapidly in the time period of the dataset. By October 27, continuum images revealed an elliptical region outlining development of the leading and trailing sunspots with the leading sunspot incompletely formed with only a partial umbra and penumbra. Hα images showed the active region enveloped in arc filaments stretching from the leading sunspot to the following sunspot.

The large, leading sunspot was fully formed by October 28 in continuum images with the following spot continuing to grow via the accumulation of pores throughout October 28 and 29. The leading sunspot also became visible in Hα images on October 28 while the following sunspot was only partially visible on October 28 and 29. Positive and negative magnetic field areas grew at similar rates during the three days with the magnetic field areas increasing in size by an order of magnitude (~$10^9$ to ~$10^{10}$ km$^2$) between October 26 and 31.

Arc filaments for both sunspots on October 27 show both CW and CCW motion at varying rates and locations around the sunspots, and over different time intervals in Hα images. Two filamentary features around the leading sunspot show fairly regular CCW motion from ~14:15 UT to ~16:50 UT at rates of 5.23° hr$^{-1}$ (0.90 km s$^{-1}$) and 1.46° hr$^{-1}$ (0.31 km s$^{-1}$). Another filamentary feature, at ~315°, moved CW at 1.43° hr$^{-1}$ (0.30 km s$^{-1}$).

Filamentary mass motion around the following sunspot on October 27 also showed CW and CCW motions. Two filamentary fragments, measured at different radial distances from the umbral center, likely represent the same filament. One segment 48.8" from the umbral center moved abruptly CW from ~20:45 UT to ~22:00 UT at 3.43° hr$^{-1}$ (0.59 km s$^{-1}$) while the segment measured at 59.6" showed irregular CW motion from ~18:15 UT to ~22:00 UT at average rates of 1.63° hr$^{-1}$ (0.34 km s$^{-1}$). A filament at a different location around the following sunspot moved consistently CCW from ~15:30 UT to ~23:00 UT at 2.22° hr$^{-1}$.

Filaments surrounding the leading sunspot dominate the motion seen on October 28. One filament, moving at a CCW rate of 4.35° hr$^{-1}$ (0.91 km s$^{-1}$), only shows motion from ~14:30 UT to ~16:30 UT. Three different filaments, however, show movement from ~14:30 UT to ~22:30 UT at average rates of 0.66° hr$^{-1}$ (0.11 km s$^{-1}$: CCW), 0.97° hr$^{-1}$ (0.17 km s$^{-1}$: CW), and 1.23° hr$^{-1}$ (0.21 km s$^{-1}$: CW).

Two individual pores trace CW and CCW rotational motion in continuum images on October 29 as they move away from the leading sunspot. Figure 6 shows a time-slice plot of the two pores at ~170° where the top pore is moving CW at an angular rate of 0.57° hr$^{-1}$ (0.12 km s$^{-1}$) while the bottom pore moves CCW at a rate of 0.38° hr$^{-1}$ (0.07 km s$^{-1}$). Two pores around the following sunspot were moving CW at rates of 2.15-2.31° hr$^{-1}$ (0.31-0.37 km s$^{-1}$) while a third pore was moving CCW at 0.57° hr$^{-1}$ (0.12 km s$^{-1}$).

Measurable filamentary motion on October 29 was restricted to the following sunspot, although some short-duration motion to the southwest of the leading sunspot was seen.

Filamentary motion occurred primarily to the southeast of the following sunspot and was visible throughout the day's images. One filament, measured at average distances of 38.8" and 48.8" from the umbral center, displays motion at rates of 1.16-2.16° hr$^{-1}$ (0.16-0.37 km s$^{-1}$: CCW). Another filament ~60° away rotated CW at 1.06° hr$^{-1}$ (0.22 km s$^{-1}$).

The photospheric pore motions and the filament motions measured in this young active region are not related, but are typical of both photospheric and chromospheric activity in a growing bi-polar sunspot pair and active region.

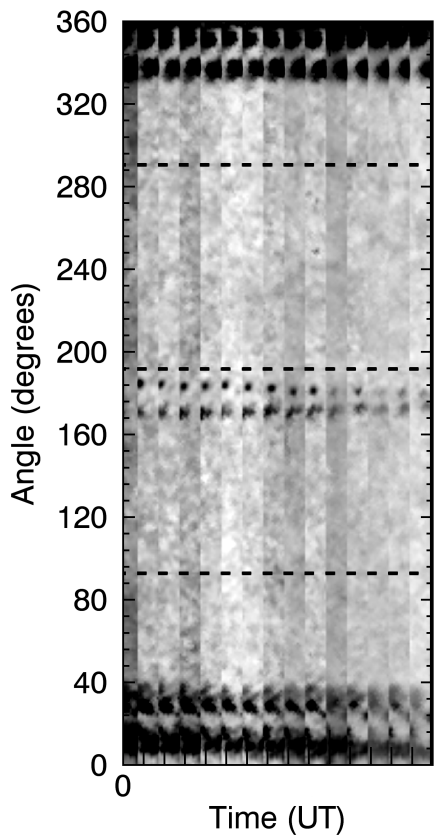

Figure 6. Time-slice plot of the open photosphere to the west of the leading sunspot in NOAA AR 10488 on October 29, 2003. The time span of the data ranges from ~14:14 UT to ~22:30 UT. Two pores are approaching each other at ~180° and appear to be in the process of merging. The black horizontal dashed lines are artifacts of the data reduction process.

4. Analysis and Conclusions

From the work of Zhang et al. (2008) and this work, it appears that countervailing rotational motions are possible around the same sunspot at the same time. Zhang et al. (2008) report CCW rotation on October 29 for both sunspots in NOAA AR 10486 while this work identified an extended umbral segment to the west/southwest of the leading sunspot, embedded in penumbra, that exhibits CW rotation. The small region of umbral CW motion seen in this work has a negative magnetic polarity that directly abuts the region of large positive polarity for the leading sunspot, as seen in SOHO magnetograms.

Another interesting issue is the lack of large-scale, complete rotational penumbral motion around a sunspot umbra. Previous works have identified various small features around a sunspot that exhibit rotation (Brown et al. 2003; Yan & Qu 2007; Min & Chae 2009; Kazachenko et al. 2009, 2010; Yan et al. 2009; Zhang et al. 2007). However, there has yet to be a report of full penumbral rotation around a sunspot center. The extent of mass entrainment seen in the following sunspot in NOAA AR 10486 on October 29 is significant and hints at a region with magnetic fields of sufficient strength to entrain the surrounding plasma in the photosphere. The lack of homogeneous 360° rotation around a sunspot suggests variable magnetic field strengths, which may be detectable and measurable in extreme ultraviolet/X-ray coronal images.

The region of mass motion in the following sunspot in NOAA AR 10486 on October 29 displays linear to semi-linear motion during, and after, the X10 flare. No major rotational acceleration or deceleration resulted from the occurrence of the flare. Kazachenko et al. (2010) suggests that rotation in AR 10486 on October 28 provided a negligible component to the helicity and energy budget in the region where flaring occurred. If this condition persisted into October 29, then that may be why there was little or no rotational response during or after the X10 flare.

Lastly, NOAA AR 10486 displays some indications of mass coupling of the photosphere and chromosphere on October 29, primarily before the X10 flare. While the rotation of regions of the photosphere is prominent (Figure 3) in the following sunspot, CCW chromospheric motion is much smaller in magnitude and more episodic. Some chromospheric mass also appears to have been displaced as a result of the X10 flare (Figure 7). This chromospheric motion is time-coincident with the resulting proton storm as seen in SOHO EIT images of the same time period. The abrupt filament movements in close association with the X10 flare could have been caused by an Hα Moreton wave or a magnetosonic shock wave propagating through the lower regions of the solar atmosphere (Shen & Liu 2012; Asai et al. 2012).

One suggestion for the observed photospheric-chromospheric coupling is the presence of magnetic field lines of varying inclinations from the vertical that are preferentially stronger in the penumbral regions of mass motion compared to the field lines originating from other locations in the penumbra. While it is well known that sunspot magnetic field strength decreases with increasing radial distance from the sunspot center (Bhatnagar & Livingston 2005), this result suggests greater variations in magnetic field strength as a function of position around the entire penumbra.

Hα images in this work sample the core of the feature at 656.28 nm (with a bandwidth of ~0.080 Å). Radiative transfer and model atmosphere calculations for the disk chromosphere show that the core of the Hα line samples the lower- to middle-portions of the average chromosphere (1600-3000 km: Wilson and White, 1966; 1200-1700 km: Vernazza et al., 1981; 1107-2100 km: Qu and Zu, 2002; 1900 km: Balasubramaniam et al., 2007; 500-1500 km: Leenaarts et al., 2012). Hence, we estimate that the measurements here represent the average photosphere and chromosphere and are, at best, separated by a few thousand kilometers.

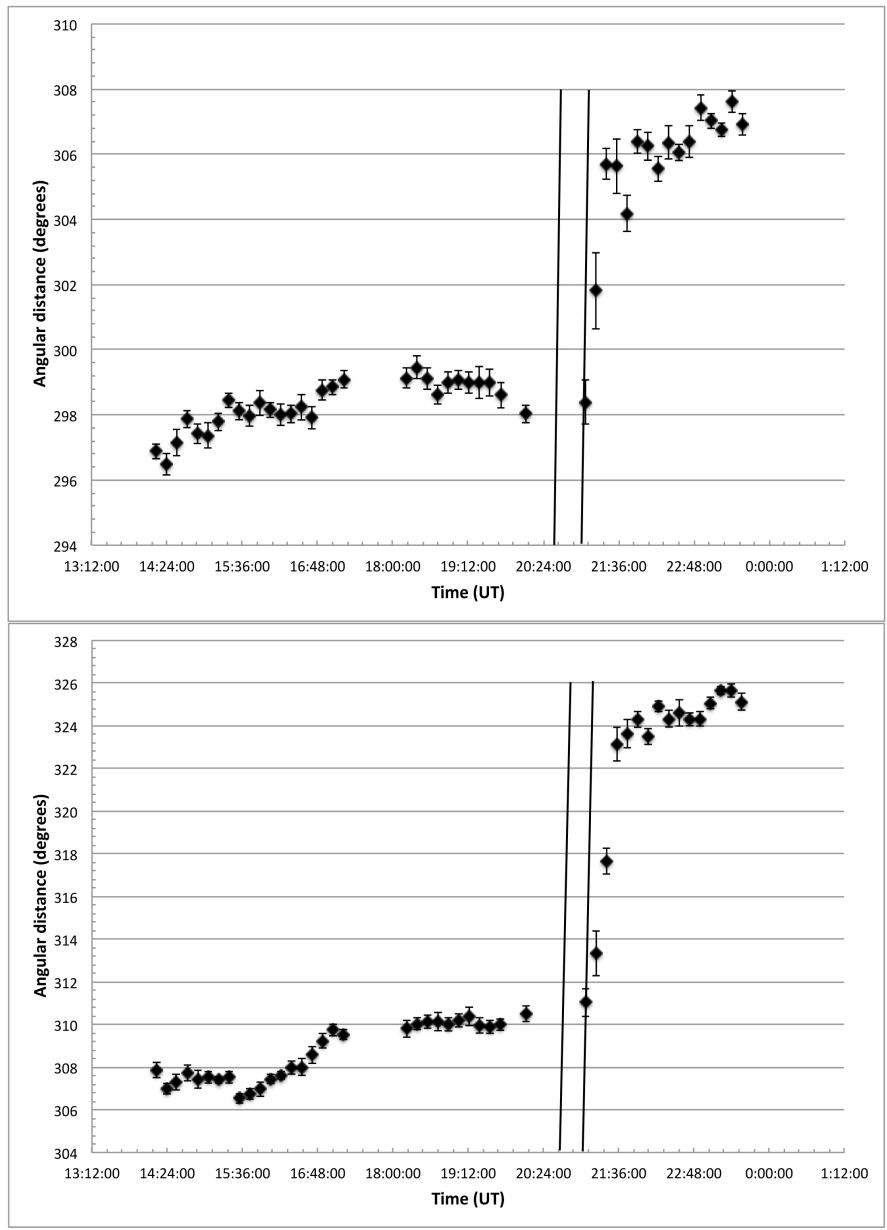

Figure 7. Episodic counterclockwise motion of chromospheric mass of two features in the following sunspot of NOAA AR 10486 during and immediately following an X10 flare, which is indicated by the offset of two chromospheric features on October 29, 2003, at ~20:45 UT. The X10 flare began at 20:37 UT and ended at 21:01 UT, which is indicated by the vertical lines in both figures. The location of this mass movement is near, or adjacent to, the mass motion in the photosphere.

We find it important to address the possible application of non-linear force-free (NLFF) modeling methods to understanding the rotational coupling in the photosphere and chromosphere. Non-linear force-free models (see e.g. Wiegelmann & Sakurai 2012; Wiegelmann et al. 2012) are still in their infancy. The vertical resolution/granularity of the extrapolation is un-estimated currently (perhaps a few thousand kilometers) and never verified. It's not a self-consistent problem where one can fit the observed magnetic fields to vertical structures.

Modeling of the magnetic field requires vector magnetic measures that are limited to one layer, namely the photosphere, which is also its lower bound. We are still unsure how surface errors of the measured longitudinal field (up to 20% for weak fields, e.g., a 5 G error in a 20 G measurement) and transverse fields (~50%, e.g., a 100 G error in a 200 G measurement) translate to an error in the vertical direction as vertical fields are assumed to have an exponential drop in the modeling processes. Measuring the change in magnetic field with height in the solar atmosphere (i.e., *dB/dz*) also requires accurate measures of gas pressure and magnetic pressure variations with altitude. Consistent measurements for photospheric magnetic fields are now common (Petrie, 2012), but chromospheric magnetic field measurements are still lacking. Hence, using extrapolation methods is unproven to explain the twisting of sunspots in two layers.

We plan to further analyze similar data sets to understand any coupling relationships between the photosphere and chromosphere, using both ground-based ISOON and Solar Dynamics Observatory (SDO) data.

We are grateful for the use of the ISOON data. PSH and SS were supported by UND for this work. KSB was supported by the AFOSR and AFRL in this work.